\documentclass[aps,pre,twocolumn,superscriptaddress,showpacs,showkeys ]{revtex4-1}

\usepackage{graphicx}                   
\usepackage{hyperref}                   
\usepackage{amsmath,amssymb}            
\usepackage{epstopdf}
\usepackage{subcaption}					

\usepackage{color}
\definecolor{orange}{rgb}{1,0.5,0}
\definecolor{brown}{rgb}{0.65, 0.16, 0.16}
\definecolor{phlox}{rgb}{0.87, 0.0, 1.0}

\graphicspath{{figs/}}					

\begin{document}

\title{Power spectrum of rare events in two-dimensional BTW model,\\
violation of $1/f$ noise hypothesis}

\author{Z. Moghadam}
\affiliation{Department of Physics, University of Mohaghegh Ardabili, P.O. Box 179, Ardabil, Iran}
\email{zahramoghadam.physics@gmail.com }

\author{M. N. Najafi$^*$}
\affiliation{Department of Physics, University of Mohaghegh Ardabili, P.O. Box 179, Ardabil, Iran}
\email{morteza.nattagh@gmail.com}

\author{A. Saber}
\affiliation{Department of Physics, University of Mohaghegh Ardabili, P.O. Box 179, Ardabil, Iran}
\email{ahad.saber@gmail.com }

\author{Z. Ebadi}
\affiliation{Department of Physics, University of Mohaghegh Ardabili, P.O. Box 179, Ardabil, Iran}
\email{z.ebadii@uma.ac.ir}

\begin{abstract}
One of the primitive aims of the two-dimensional BTW model had been to explain the $1/f^{\alpha}$ noise which is widely seen in the natural systems. In this paper we study some time signals, namely the activity inside an avalanche ($x(t)$), the avalanches sizes ($s(T)$) and the rare events waiting time (REWT) ($\tau(n)$ as a new type of noise). The latter is expected to be important in predicting the period and also the vastness of the upcoming large scale events in a sequence of self-organized natural events. Especially we report some exponential anti-correlation behaviors for $s(T)$ and $\tau(n)$ which are finite-size effects. Two characteristic time scales $\delta T_{s}$ and $\delta T_{\tau}$ emerge in our analysis. It is proposed that the power spectrum of $s(T)$ and $\tau(n)$ behave like $\left( b_{s,\tau}(L)^2+\omega^2\right)^{-\frac{1}{2}}$, in which $b_{s}$ and $b_{\tau}$ are some $L$-dependent parameters and $\omega$ is the angular frequency. The $1/f$ noise is therefore obtained in the limit $\omega\gg b_{s,\tau}$. $b_{s}$ and $b_{\tau}$ decrease also in a power-law fashion with the system size $L$, which signals the fact that in the thermodynamic limit the power spectrum tends to the Dirac delta function.
\end{abstract}


\maketitle

\section{Introduction}
Many natural processes are explainable in terms of some local simple rules and show degrees of self-organized criticality (SOC)~\cite{bak1989earthquakes,sornette1989self,charbonneau2001avalanche,clar1999self,kaulakys2009modeling,zhang20001,Dhar1990Self}, most of which show $1/f^{\alpha}$ power-law decay in the tails of their power spectra of time signals. The term flick noise refers to the phenomenon that a signal $s(t)$ fluctuates with a power spectrum $PS_s(f)=1/f^{\alpha}$ at very low frequencies. Since the exponent $\alpha$ is often close to $1$, flick noise is also called $1/f$ noise~\cite{zhang20001}. Examples are electrical noise~\cite{carreras2004evidence}, solar flares~\cite{charbonneau2001avalanche}, stock market price variations, rain~\cite{Dhar1990Self} and earthquake~\cite{bak1989earthquakes}. The destructive features of the large scale events in the natural disasters motivates one to model the harmony of the occurrence of them which helps to predict their behaviors~\cite{alexander1993natural}. Due to the random nature of the problem, statistics and probability analysis are extensively deployed as basic mathematic tools to analyze rare events based on historical data. However, it is difficult to acquire accurate and adequate data in practice as very limited information can be recorded. In this regard the theoretical modeling and the statistical analysis on the corresponding rare events yields some valuable information concerning their sequence pattern and also their intensities. Despite of random nature of the SOC systems, their power-law harmony restricts strongly their spatial and temporal correlations and other behaviors, giving the chance to make some predictions on the future of the sequence~\cite{Dhar1990Self}.\\ 
The sandpile model as a prototype of self-organized critical systems is a dissipative system which is slowly driven and displays irregular bursts of scale free large activity named as avalanche. This model was designed firstly by Bak \textit{et al.} to explain the $1/f^{\alpha}$ noise in natural phenomena~\cite{BTW1988Self}. These models show critical behavior without fine tuning of any external parameter. The examples of the systems that this idea have had impact are ecology~\cite{halley1996ecology}, cognitive processes~\cite{beggs2003neuronal}, magnetic systems~\cite{bertotti2006science}, superconductors~\cite{field1995superconducting}, mechanics \cite{petri1994experimental,salminen2002acoustic} and magnetosphere~\cite{chang2003complexity,chang1999self}. Studying the BTW sandpile model as the first example of such systems is very helpful to find the temporal and spatial structure of correlations, which is vital for predicting the future of a sequence of avalanches. \\
Soon after the work of BTW, many authors investigated the correlations and also the power spectrum of the sandpile models and many estimations of the decay exponents were obtained, ranging from Lorentzian $\alpha=2$~\cite{Jensen1989critical} to more exact exponent $\alpha=1.59$~\cite{Laurson2005power}. Two types of correlations should be distinguished: within a single avalanche, and over time scales greater than the interval between successive particle additions, which measures correlations between avalanches. The $\alpha$ exponent has been obtained for the relaxation events inside an avalanche in the mentioned references, and less attention has been paid to the statistical correlations between distinct avalanches~\cite{baiesi2006realistic,hwa1992avalanches,kutnjak1996temporal}. The importance of the latter case can be understood in rare events statistics which is expected to have connections with natural self-organized rare events such as earthquake and making predictions. The precise determination of correlation/anti-correlation and their decay times is also important, e.g. in estimating the time required for rare events to become nearly independent.\\
Many attempts have been made to find the structure of temporal and spatial correlations/anticorrelations in sanpile models~\cite{hwa1992avalanches,ali1995self,kutnjak1996temporal,ahmed2010avalanche,majumdar1991height,dhar1990abelian,lubeck2000crossover}. Anticorrelations in sandpiles are caused by discharge effects, meaning that after a large scale event occurs, the system loses dramatically its energy content and needs some time to re-obtain the necessary energy to trigger another large event. In this paper we consider 2D BTW sandpile model and analyze three time scales: $x(t)$ which is the number of unstable sites in the internal time $t$, $s(T)$ which is $T$th avalanche and $\tau(n)$ which is the waiting time between $n$th and $(n+1)$th avalanches. We observe that a time scale emerge for $s(T)$ and $\tau(n)$ which limits the domain of the temporal anti-correlations, leaving the only possibility that the corresponding power spectrum behaves like $\left(b_{s,\tau}^2+\omega\right)^{-\frac{1}{2}}$ in which $b_s\equiv 1/\delta T_s$, $b_{\tau}\equiv 1/\delta T_{\tau}$ and $1/\delta T_s$ and $1/\delta T_{\tau}$ are the time scales for $s(T)$ and $\tau(n)$ respectively. The $x(t)$ noise however show scale-free behaviors with two distinct intervals, having its roots in the multi-fractality of the avalanches. The nearly $1/f^{1.6}$ behavior is observed for this noise in accordance with the previous results.\\

The paper has been organized as follows: In the following section, we explain the method and define the functions. In SEC~\ref{results} we present the results for the three noises. We end the paper by a conclusion.

\section{The construction of the problem}
\label{sec:model}
The main ingredients of avalanche propagation in natural phenomena (as well as the sandpile models) are a slow driving, a local threshold (or non-linearity) for the dynamics and a dissipation mechanism.

Before going into the details, let us first briefly introduce the standard BTW model on a regular $d$-dimensional hypercubic lattice \cite{Bak1987Self}. Let each site of the lattice $i$ has an integer height (energy) $ E_i \ge 1$. At initial state, one can set randomly the height of each site in which $ E_i \le E_c $. $ E_c $ is the threshold height equal to the number of nearest neighbors of each site (e.g, for hypercubic lattice $ E_c=2d $). At each time step a grain is added on a randomly chosen site ($ E_i \to E_i+1 $). If the height of this site exceeds $E_c$, a toppling occurs: $ E_i\to E_i+\Delta_{i,j} $ in which $\Delta_{i,j}=-E_c$ if $i=j$, $\Delta_{i,j}=1$ if $i$ and $j$ are neighbors and zero otherwise. A toppling may cause the nearest-neighbor sites to become unstable (have hight higher than $E_c$) and topple in their own turn and so on, until the entire lattice sites are below the critical threshold (stable state). The total process which starts by a local perturbation (making the first site unstable) until reaching another unstable configuration is called an avalanche. The model is conservative and the energy is dissipated only from the boundary sites. The properties of the model in $d=2$ has been investigated extensively and well understood in the literature \cite{Dhar2006Theoretical}, as well as $d=3$ \cite{dashti2015statistical,Lubeck1997BTW,Ktitarev2000Scaling}.\\
We have two kinds of time, namely the \textit{internal time} (denoted by $t$) and the \textit{external time} (denoted by $T$). For defining these times, we have partitioned the total number of topplings in an avalanche into some internal time steps. One unit of internal time is defined as a \textit{$L^2$ search for unstable sites}. Therefore one can define the sequence $\left\lbrace x(t) \right\rbrace_{t=1}^{t_{\text{max}}}$ in which $x(t)$ is the number of topplings at time $t$. The sum of activities in an avalanche $\sum_{t=1}^{t_{\text{max}}(T)}x(t)$ ( in which $t_{\text{max}}(T)\equiv$ total internal time needed for the $T$th avalanche to be ended) equals the total size of that avalanche $s(T)$ ($\equiv  \#$ toppling events in the $T$th avalanche) which can be considered as another noise. In fact $x(t)$ records the number of topplings (local relaxation events) taking place in the sandpile during each parallel update of the whole lattice, one such update defining the unit of internal time, whereas $s(T)$ records the avalanche size in $T$th avalanche.\\
For the internal time signal ($x(t)$) of avalanche one can present the following scaling argument to obtain the $\alpha$ exponent. The power spectrum is defined as 
\begin{equation}
PS_x(f)\equiv \left\langle \left| \frac{1}{t_0}\int_0^{t_0} dt x(t,t_0)\exp\left[-2i\pi ft\right] \right|^2 \right\rangle 
\end{equation}
in which $\left\langle \right\rangle$ show ensemble averaging and $t_0$ is the time duration of the avalanche and $x(t,t_0)$ is time series conditioned to be terminated in $t=t_0$. Note that in this ensemble averaging, the duration of $t_0$ is also sample-dependent. To facilitate the notation we use the quantity $PS(f,t_0)$ which is the power spectrum of avalanches with fixed duration $t_0$, for which:
\begin{equation}
\begin{split}
PS_x(f|t_0)\equiv &\frac{1}{t_0^2}\iint_0^{t_0}dtdt' \exp\left[-2i\pi f(t-t')\right] \\
&\times \left\langle x(t,t_0)x^*(t',t_0)\right\rangle
\end{split}
\end{equation}
The total power spectrum is readily obtained by a second averaging over the avalanche sizes, i.e. $PS_x(f)=\int P(s)PS_x(f|s)ds$ in which $P(s)$ is the probability density function of $s$. Knowing that the avalanche size scales with the its time duration via the relation $s\sim t_0^{\gamma_{st_0}}$, and using the facts that $PS_x(f|s)=s^2F_1(f^{\gamma_{st_0}}s)$ and $x(t,t_0)=t_0^{\gamma_{st_0}-1}F(t/t_0)$~\cite{Laurson2005power} and also $P(s)\sim s^{-\tau_s}$, one obtains:
\begin{equation}
PS_x(f)= f^{-\gamma_{st_0}(3-\tau_s)}\int_0^{s^*f^{\gamma_{st_0}}}dxx^{2-\tau_s}F_1(x).
\end{equation}
in which $s^*$ is the size cut-off of avalanches which is determined by finite size of the system. If we let $s^*\rightarrow \infty$ the integral equals a constant and one obtains $\alpha=\gamma_{st_0}(3-\tau_s)$.\\
The similar arguments also hold for the time series $s(T)$ as a signal. Here the power spectrum is obtained without ensemble averaging, by the following relation:
\begin{equation}
PS_s(f)\equiv \lim_{T_0\rightarrow\infty}\left| \frac{1}{T_0}\int_0^{T_0} dT s(T)\exp\left[-2i\pi fT\right] \right|^2 
\end{equation}
In calculating this quantity, we can begin by one initial configuration and let it to reach the steady state, after which the above quantity is obtained up to some large time $T_0$.\\
\begin{figure*}
	\begin{subfigure}{0.49\textwidth}\includegraphics[width=\textwidth]{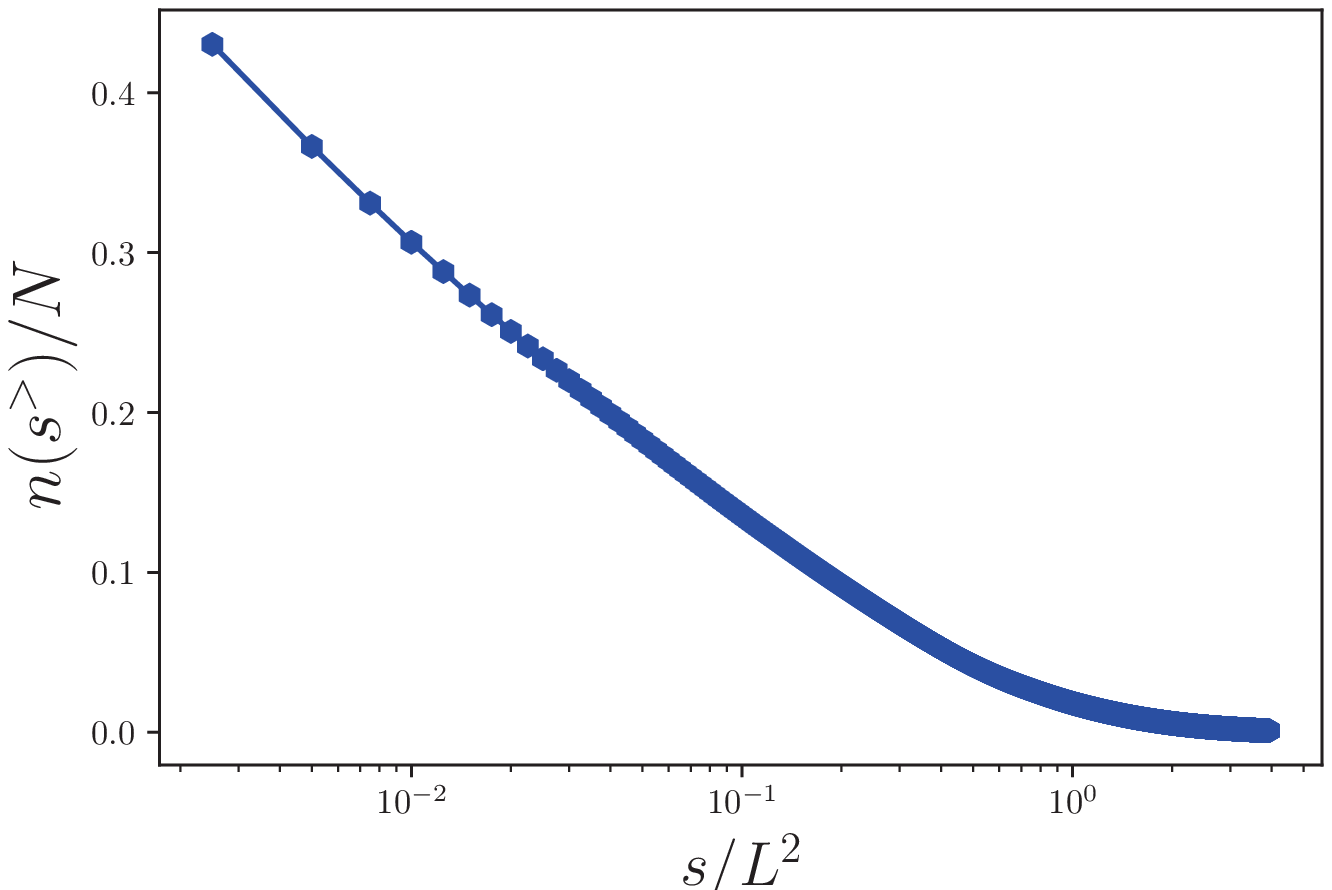}
		\caption{}
		\label{fig:treshold}
	\end{subfigure}
	\centering
	\begin{subfigure}{0.49\textwidth}\includegraphics[width=\textwidth]{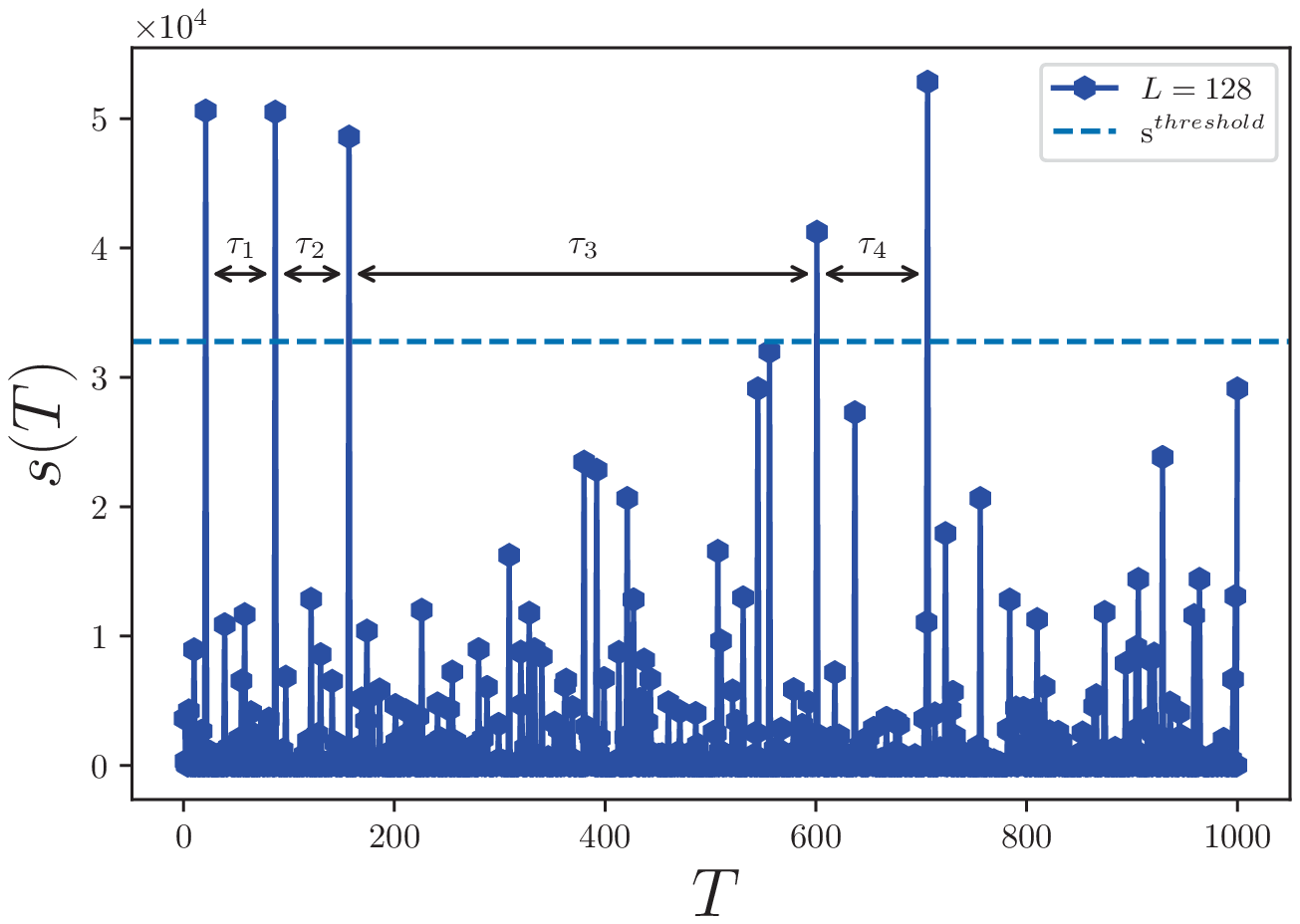}
		\caption{}
		\label{fig:tau}
	\end{subfigure}
	\caption{(Color online): (a) Semi-log plot of $n(s^>)/N$ in terms of $s/L^2$. The graph becomes slow-varying for $s\gtrsim L^2$. (b) a time signal sequence $s(T)$ for $L=128$. The threshold has been considered to be $s^{\text{threshold}}=2L^2$. The first four rare event waiting times have been shown by $\tau_1$, $\tau_2$, $\tau_3$ and $\tau_4$.}
	\label{schematics}
\end{figure*}
Another quantity which can be seen as a time series is the \textit{rare events waiting time} (REWT). Apparently we have freedom to define a rare (large scale) event. Fixing a proper definition of the rare events needs some statistical analysis to minimize the sensibility of the results to the definition. In this paper we have defined a rare avalanche as an avalanche whose size is more than twice the lattice size, i.e. $s^{\text{threshold}}=2L^2$. In fact we have examined various rates for this threshold value (for defining the rare event) and have done simulations for $s^{\text{threshold}}=L^2,1.5L^2,2L^2,3L^2$ and $5L^5$ and the results have not been changed. In Fig.~\ref{fig:treshold} we have shown $n(s^>)/N$ in terms of $s^{\text{threshold}}/L^2$ in which $n(s^>)$ is the number of avalanches larger than threshold and $N$ is the total number of avalanches. It is seen that for $s^{\text{threshold}}\gtrsim 10^0$ $n(s^>)/N$ becomes slow-varying. The more general quantity is the waiting time between avalanches with sizes larger than $s$, i.e. $t_w^{>s}$~\cite{baiesi2006realistic}. To avoid unnecessary complications, we have fixed the threshold. After fixing the definition of large events, we define REWT, denoted by $\tau(n)$, as the time interval between two successive rare events $s(T_{n})$ and $s(T_{n+1})$. In other words if the $(n)$th rare event occurs at time $T_{n}$ (i.e. $s(T_{n})>s^{\text{threshold}}$) and the next rare event occurs at time $T_{n=1}$, then $\tau(n)\equiv T_{n+1}-T_{n}$. The REWT has been shown in Fig.~\ref{fig:tau} for a system with size $L=128$ in which the threshold has been shown by a broken line and some $\tau$'s have been shown. Therefore a time series of rare events is formed, which is denoted by $\left\lbrace \tau(n)\right\rbrace_{n=1}^{n_{\text{max}}}$ in which $n_{\text{max}}$ is the maximum of $n$ in the simulations and $T_0\equiv0$. Now the corresponding power spectrum is calculated by the following relation:
\begin{equation}
PS_n(f)\equiv \lim_{n_{\text{max}}\rightarrow\infty}\left| \frac{1}{n_{\text{max}}}\sum_{n=0}^{n_{\text{max}}} \tau(n)\exp\left[-2i\pi fn\right] \right|^2.
\end{equation}
The statistics of REWT enables us to make some statistical predictions on the large scale events, which is important in natural disasters as explained in the introduction. Especially the autocorrelation of $s(T)$ REWT is very important in the rare event analysis. These correlation demonstrate the tendency of a large scale event to have a more/less vast event in the future, i.e. answers the question whether the rare events amplify or weaken each other in a time sequence. The answer whatever it is, seems to be of vital importance in the natural self-organized critical systems, provided that the connection to the underlying theoretical model is correct. Regardless of the exact form of the autocorrelation functions, the positivity or negativity of the time-correlations of the rare events is expected to have a deep impact in the analysis of the corresponding natural phenomena. In this paper we concentrate mainly on the time signals $s(T)$ and $\tau(n)$, and also regenerate the results for $x(t)$ which has been investigated vastly in the literature (see for example~\cite{Laurson2005power}).

\section{Results}\label{results}

We consider two-dimensional BTW model on square lattice of linear sizes $L=32,64,128,256,512$ and $1024$. For calculating our desired quantities, more than $10^7$ avalanches have been taken into account for each lattice size. We start with a random height distribution $h_i\in [1,4]$ and inject the sand grains randomly through the sample. Once the system reached the steady state, the statistical observables are analyzed. \\

\begin{figure*}
	\centering
	\begin{subfigure}{0.45\textwidth}\includegraphics[width=\textwidth]{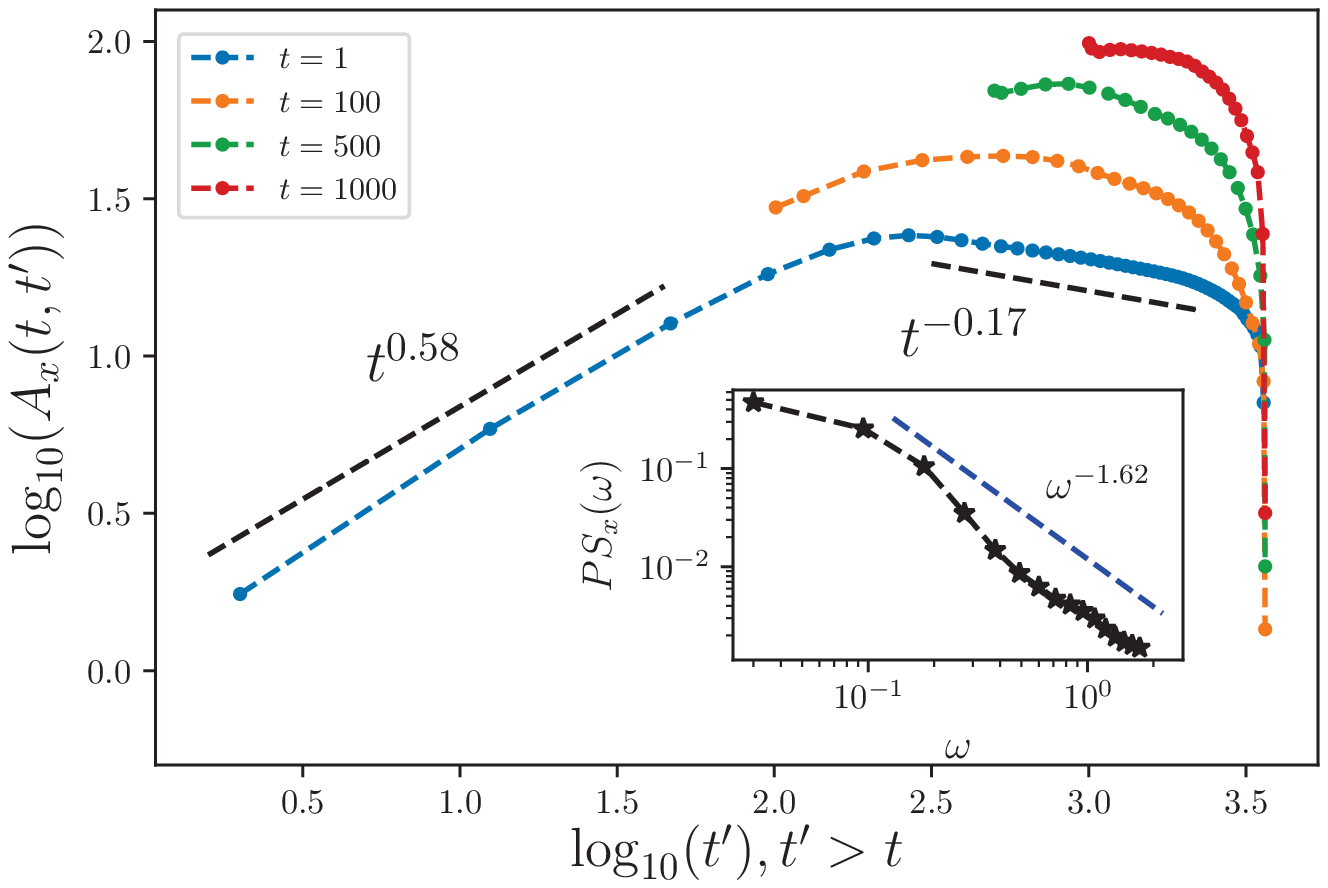}
		\caption{}
		\label{fig:x_t}
	\end{subfigure}
	\begin{subfigure}{0.45\textwidth}\includegraphics[width=\textwidth]{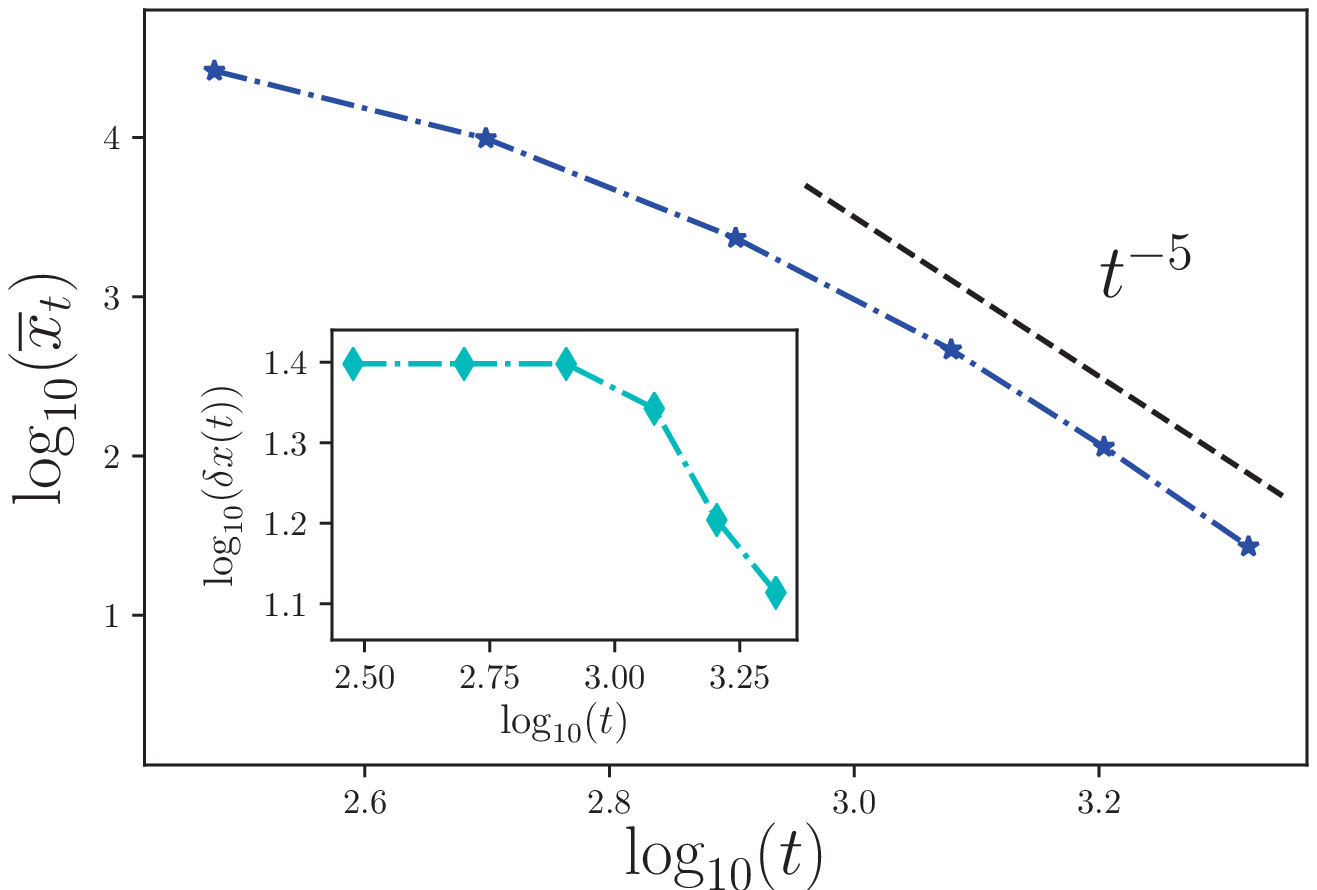}
		\caption{}
		\label{fig:xbar_t}
	\end{subfigure}
	\begin{subfigure}{0.6\textwidth}\includegraphics[width=\textwidth]{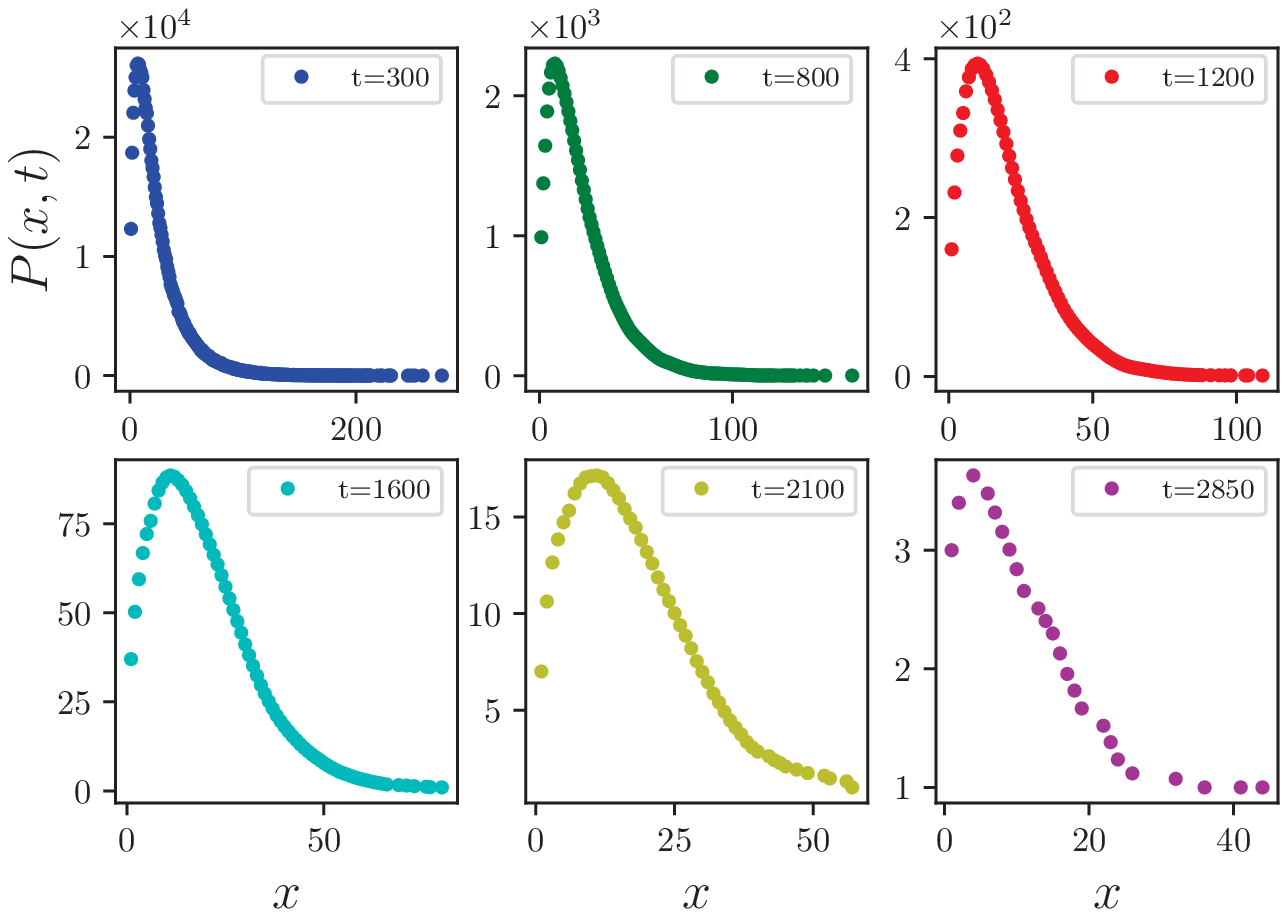}
		\caption{}
		\label{fig:P(x,t)_6}
	\end{subfigure}
	\caption{(Color online): (a) The auto-correlation function of $x(t)$. Inset: the power spectrum of $x(t)$ with the famous $\alpha\simeq 1.6$ exponent. (b) $\bar{x}_t $ and $\delta x(t)$ as functions of $t$. (c) $P(x,t)$ in terms of $x$ for various rates of $t$.}
	\label{x_t}
\end{figure*}

Figure~\ref{fig:x_t} shows the autocorrelation of the noise $x(t)$ ($A_x(t,t')\equiv\left\langle x(t)x(t')\right\rangle$) and its power spectrum ($PS_x(\omega)$) for $L=512$. For $x(t)$ the autocorrelation function $A_x(t,t')$ depends separately on both $t$ and $t'$ and not $t-t'$ (remember that it is not invariant under the transformation $t\rightarrow t+a$ and $t'\rightarrow t'+a$). This function shows multi-fractal behavior. For $t=1$, there are two distinct power-law regimes, one with $t'^{0.58\pm 0.0.02}$ and another with $t'^{-0.17\pm 0.02}$. In the inset the power-law behavior for the power spectrum is seen with the exponent $\approx 1.6$ in agreement with the previous results~\cite{Laurson2005power}. To illustrate the other features of this noise, we have investigate its other statistical properties. $\bar{x}(t)\equiv\left\langle x(t)\right\rangle$ and $\delta x(t)\equiv \sqrt{\left\langle x(t)^2\right\rangle-\left\langle x(t)\right\rangle^2 }$ have been reported in Fig.~\ref{fig:xbar_t}. For long enough times $\bar{x}(t)\sim t^{-5}$ and $\delta x(t)\sim t^{-1.0\pm 0.1}$, whereas for the small times the behaviors are different. The point at which the behavior changes is compatible with the corresponding point in Fig~\ref{fig:x_t}, and results from the multi-fractality of the avalanches. In the Fig.~\ref{fig:P(x,t)_6} we have shown the distribution function $P(x,t)$ ($\equiv$ the number of events that at internal time $t$, there are $x$ unstable sites) for various times $t$ in terms of $x$, in which $\bar{x}(t)$ is the position of their peaks and also $\delta x(t)$ is their variance. \\

\begin{figure*}
	\centering
	\begin{subfigure}{0.49\textwidth}\includegraphics[width=\textwidth]{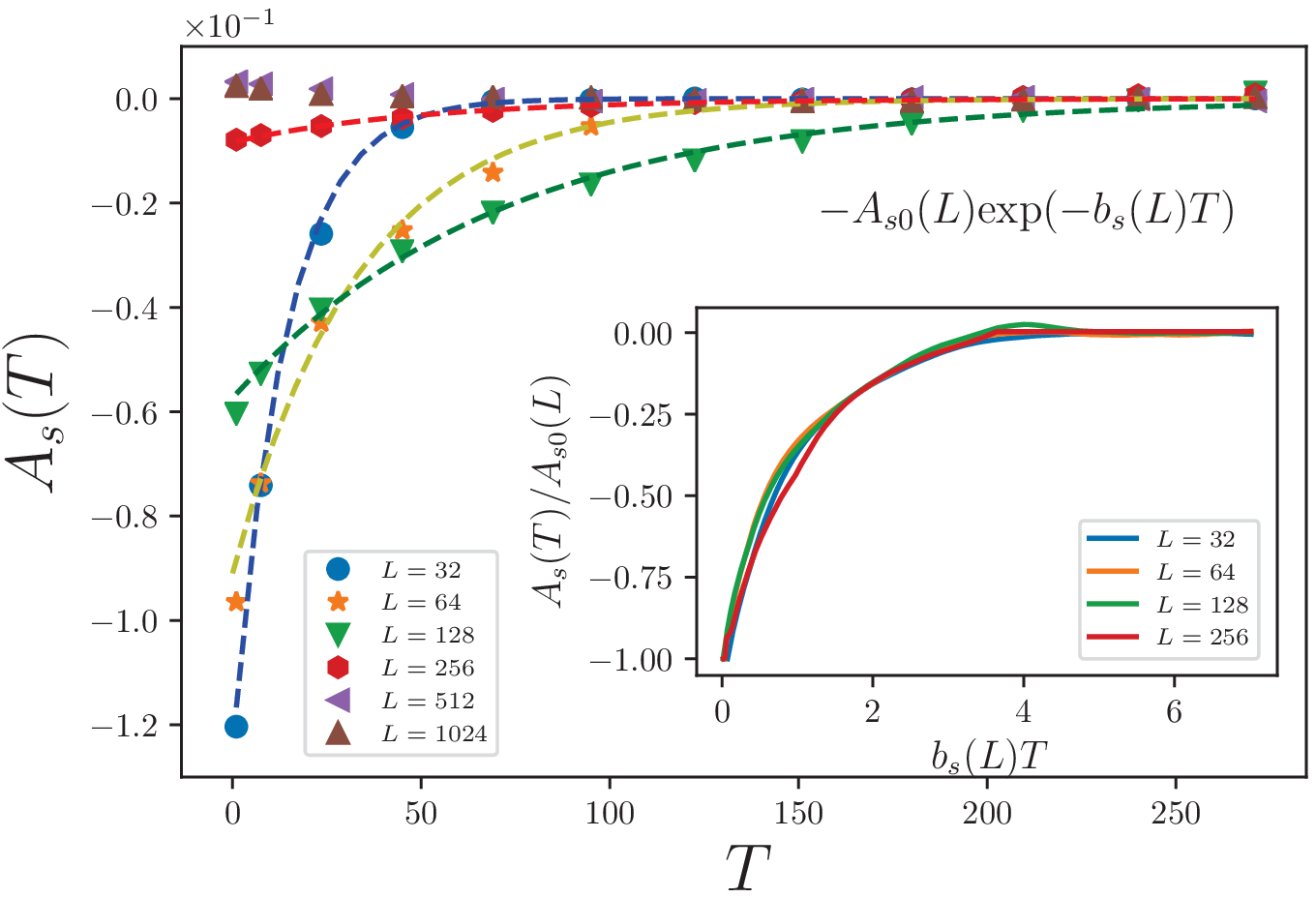}
		\caption{}
		\label{fig:A_s}
	\end{subfigure}
	\begin{subfigure}{0.49\textwidth}\includegraphics[width=\textwidth]{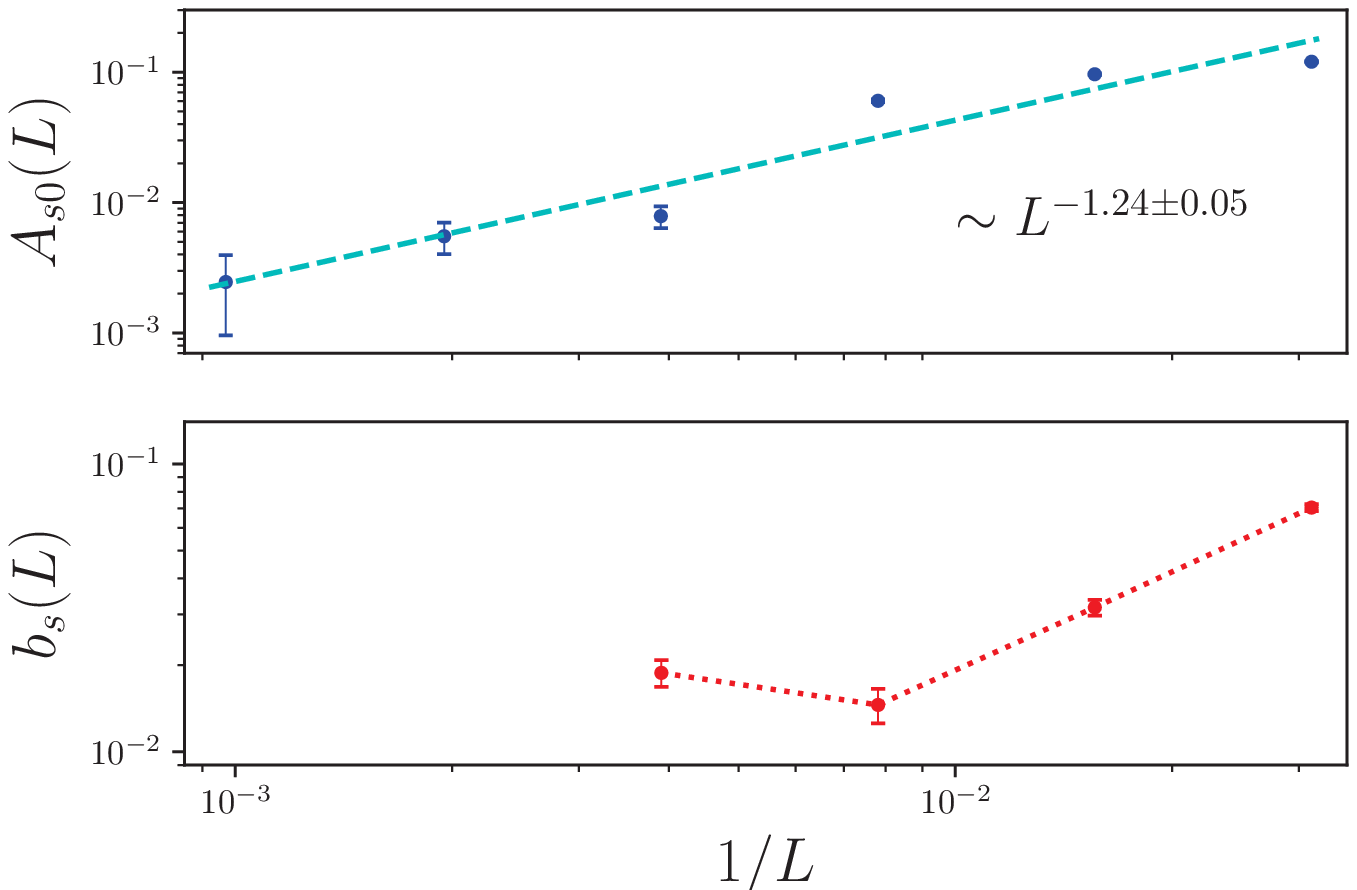}
		\caption{}
		\label{fig:A0_s,b_s}
	\end{subfigure}
	\begin{subfigure}{0.45\textwidth}\includegraphics[width=\textwidth]{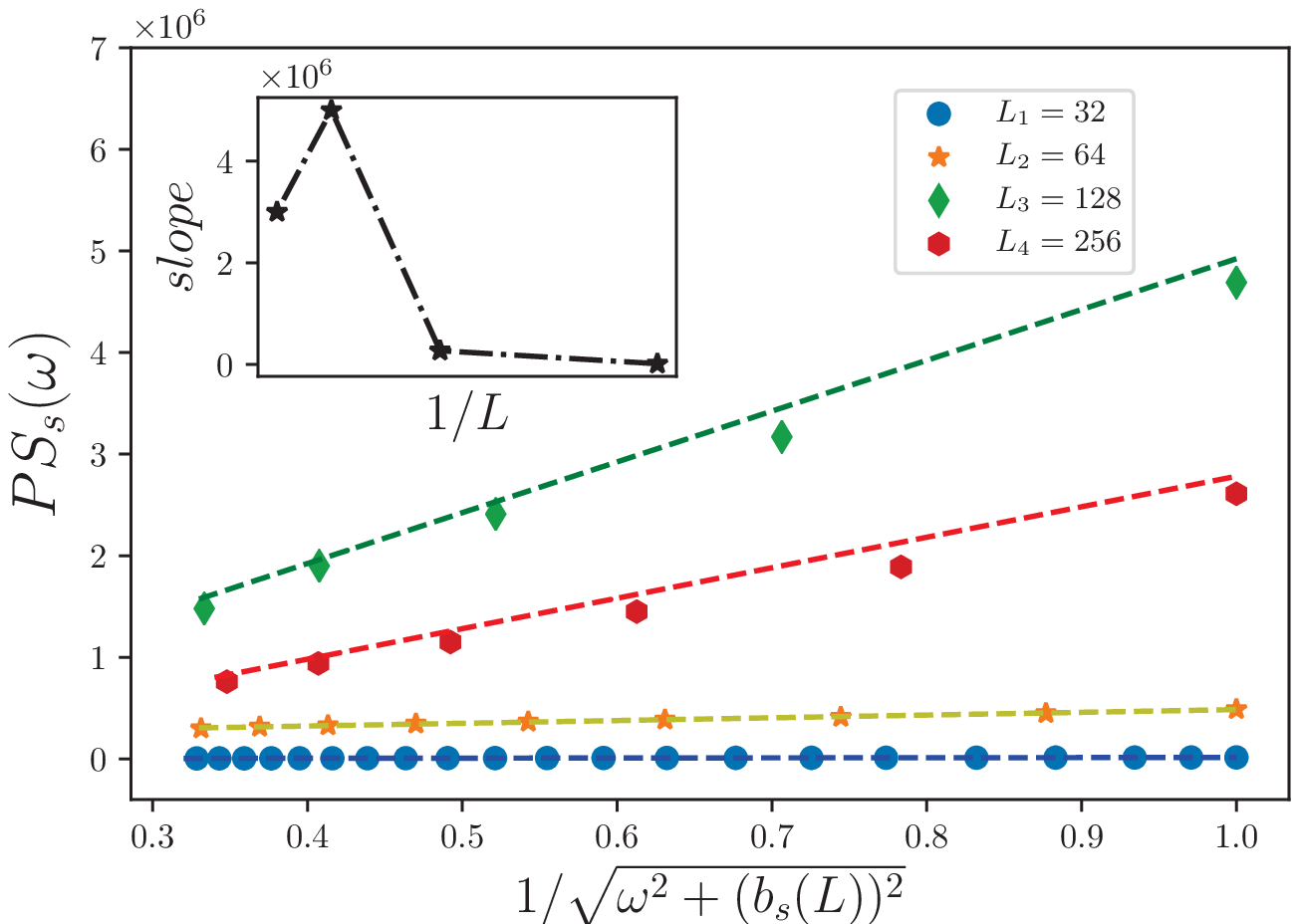}
		\caption{}
		\label{fig:PS_s,beta}
	\end{subfigure}
	\begin{subfigure}{0.45\textwidth}\includegraphics[width=\textwidth]{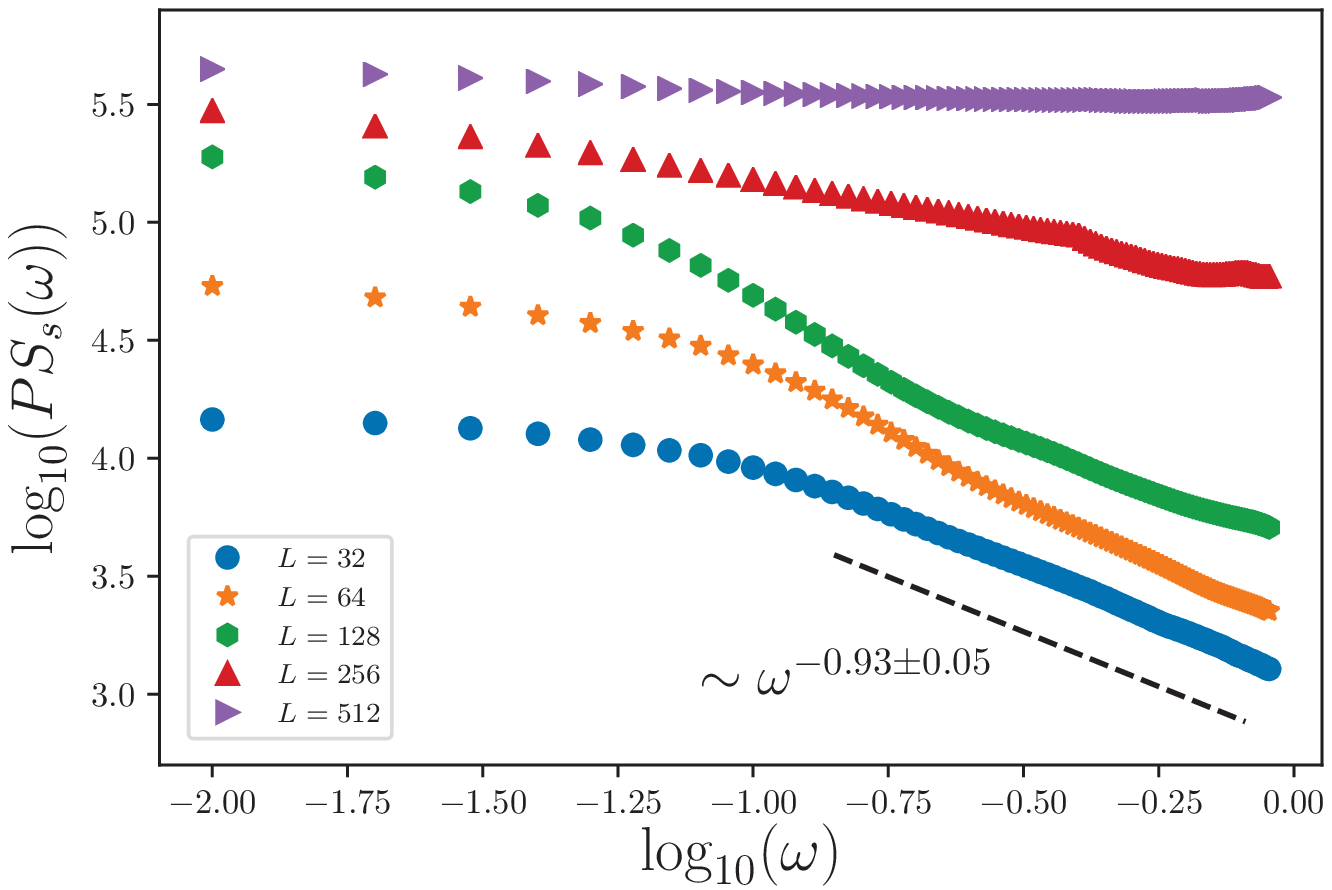}
		\caption{}
		\label{fig:PS_s,omega}
	\end{subfigure}
	\caption{(Color online): $A_s(T)$ in terms of $T$ for various rates of lattice sizes, with the corresponding fits $-A_{s0}(L)\exp\left(-b_s(L)T\right)$ (broken lines). Inset: data collapse of $A(T)/A_{s0}$ in terms of $b_sT$. (b) Log-log plot of $A_{s0}(L)$ and $b_s(L)$ in terms of $1/L$. It is seen that $A_{s0}\sim L^{-1.23\pm 0.05}$. (c) Power spectrum of $s$ ($PS_s(\omega)$) in terms of $\frac{1}{\sqrt{\omega^2+b(L)^2}}$. The slop of the fitted lines have been shown in the inset in terms of $1/L$. (d) $\log_{10}PS_s(\omega)$ in terms of $\log_{10}\omega$.}
	\label{s-statistics}
\end{figure*}

Now let us turn to the calculation of the correlation between the avalanches $s(T)$. The autocorrelation function is defined by:
\begin{equation}
f_s(T_0)\equiv \left\langle  s(T)s(T+T_0)\right\rangle_{T} -\left( \left\langle  s(T)\right\rangle_{T} \right)^2
\end{equation}
in which the $T$-average of an arbitrary statistical observable is defined by $\left\langle O\right\rangle_{T} \equiv\frac{1}{T_{\text{max}}}\sum_{T=0}^{T_{\text{max}}} O(s(T))$. Figure~\ref{fig:A_s} shows the re-scaled autocorrelation function of $s(T)$ defined by $A_s(T)\equiv \left( \left\langle  s(T)\right\rangle_{T} \right)^{-2}f_s(T)$. An interesting feature of this graph is the exponential anti-correlation of the noise $s(T)$. It is seen that $A_s(T)$ is negative for small times and grows in an exponential fashion with time approaching zero for long times. This effect is magnified for small lattice sizes. This function is properly fitted to the analytic expression $-A_{s0}\exp(-b_sT)$ for all lattice sizes considered in this paper. To show this dependence, we have shown $A_s(T)/A_{s0}$ in terms of $b_sT$ in the inset of this figure for various lattice sizes. The dependence of $b$ and $A_{s0}$ to the inverse of the lattice size $1/L$ has been shown in \ref{fig:A0_s,b_s} in a log-log plot, from which we see that $A_{s0}$ tends to zero for $L\rightarrow\infty$ with the exponent $\gamma_{A_{s0},L} \simeq \frac{5}{4}$ defined by $A_{s0}\sim L^{-\gamma_{A_{s0},L}}$. This shows that, although the correlations become more long range for larger systems, the amplitude of the anti-correlation vanishes in the thermodynamic limit and is therefore a finite size effect. The anti-correlation behavior means that when a rare event takes place at time $T$, it is not favorable for the system to have such a large scale avalanche in the approximate time interval $(T+1,T+1/b)$.\\
Having the fitted analytic form of the autocorrelation functions, one can calculate the power spectrum simply by taking the Furrier transform, which leads to the relation $|PS_s(\omega)|\sim \left( b_s(L)^2+\omega^2\right)^{-\frac{1}{2}}$. Therefore the large frequency (small time) limit of power spectrum is $PS_s(\omega)\sim \omega^{-1}$. The power spectrum of the $s(T)$ noise has been shown in Figs~\ref{fig:PS_s,beta} and \ref{fig:PS_s,omega}. The linear dependence on $1/\sqrt{b_s^2+\omega^2}$ is evident in Fig~\ref{fig:PS_s,beta} with $L$-dependent slope which is not of primary importance. The (nearly) $-1$ exponent for large $\omega$ values is seen in Fig~\ref{fig:PS_s,omega}. This is the flick noise for 2D BTW model.\\

\begin{figure*}
	\centering
	\begin{subfigure}{0.49\textwidth}\includegraphics[width=\textwidth]{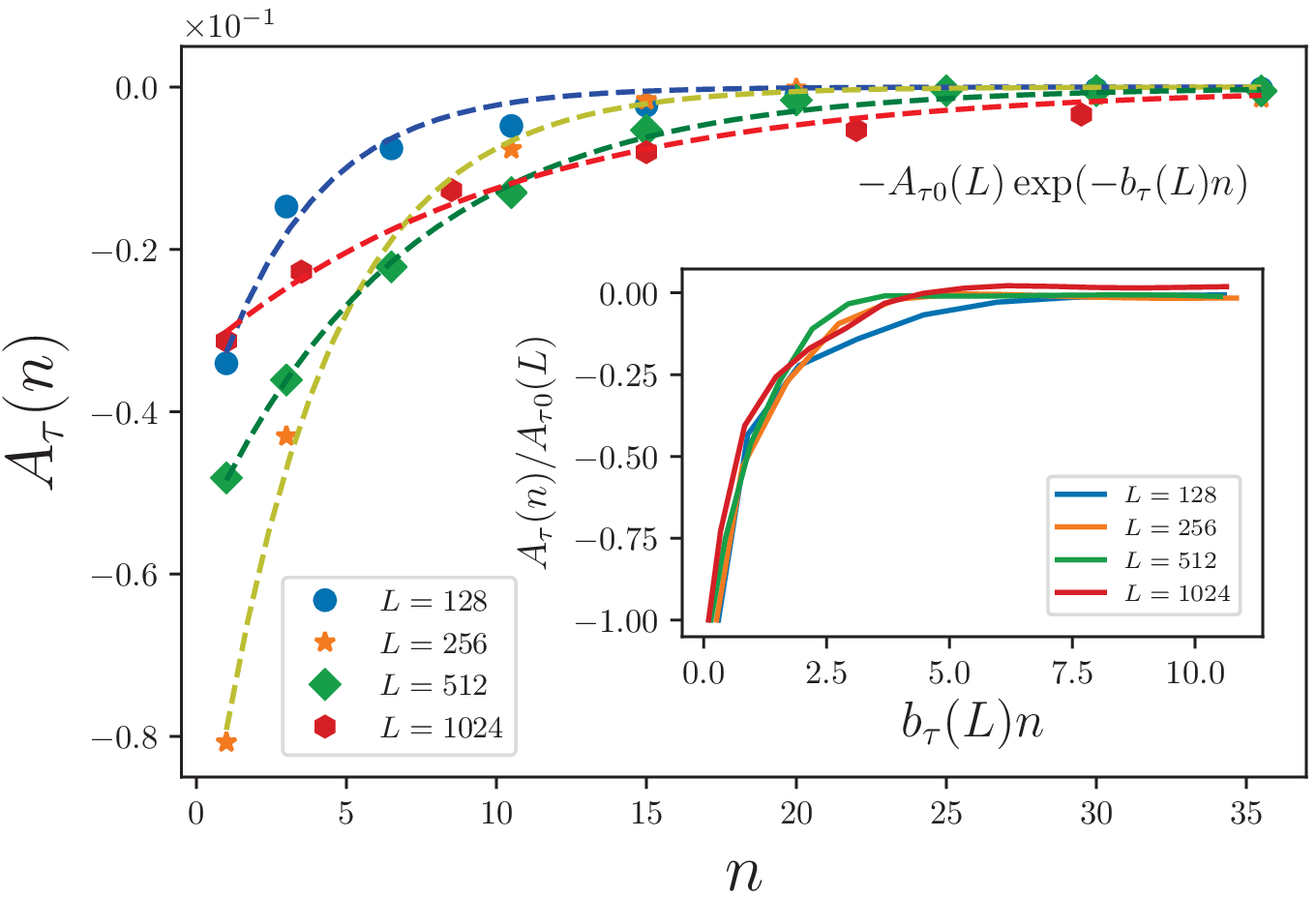}
		\caption{}
		\label{fig:A_tau}
	\end{subfigure}
	\begin{subfigure}{0.49\textwidth}\includegraphics[width=\textwidth]{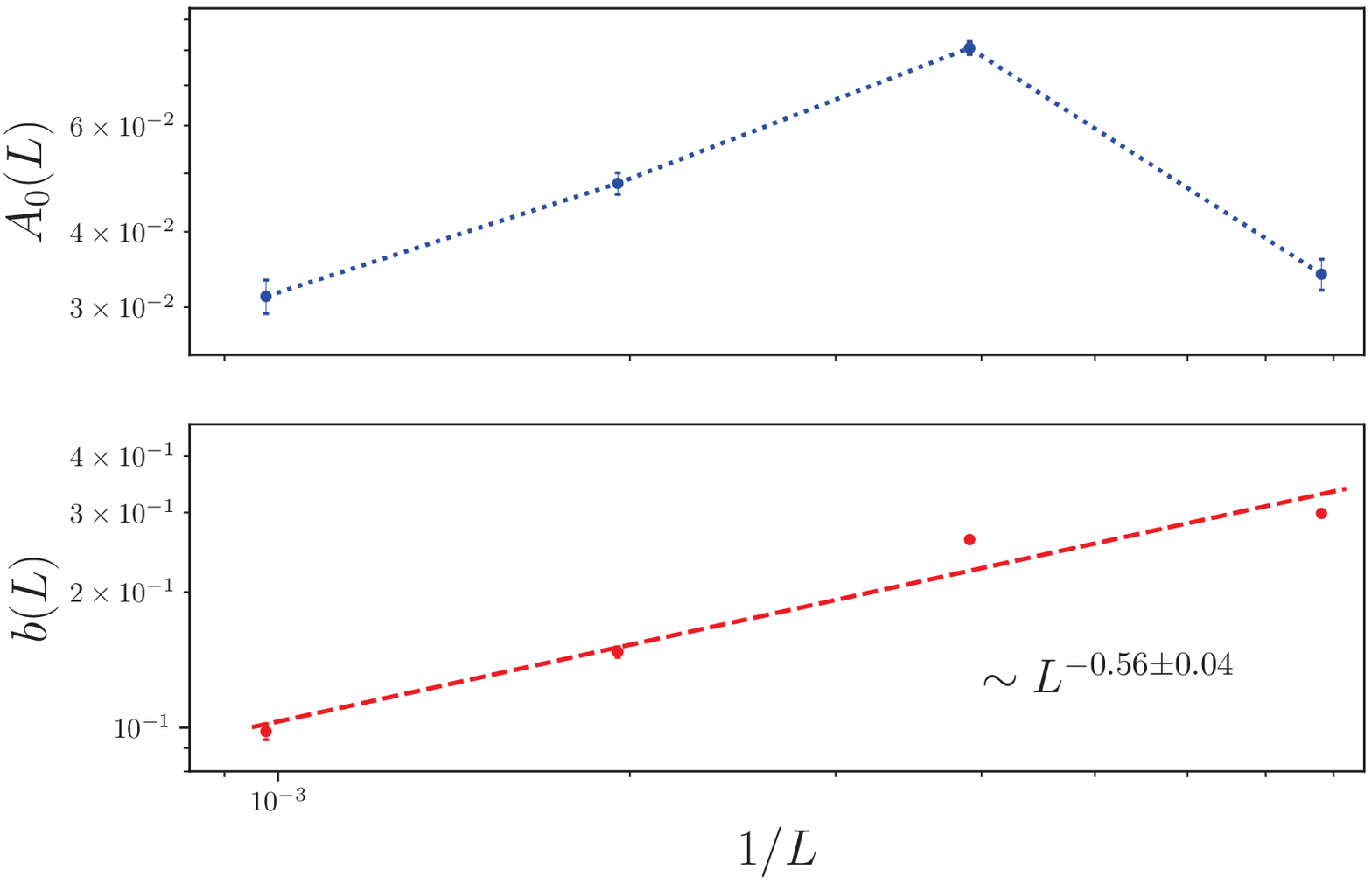}
		\caption{}
		\label{fig:A0_tau,b_tau}
	\end{subfigure}
	\begin{subfigure}{0.45\textwidth}\includegraphics[width=\textwidth]{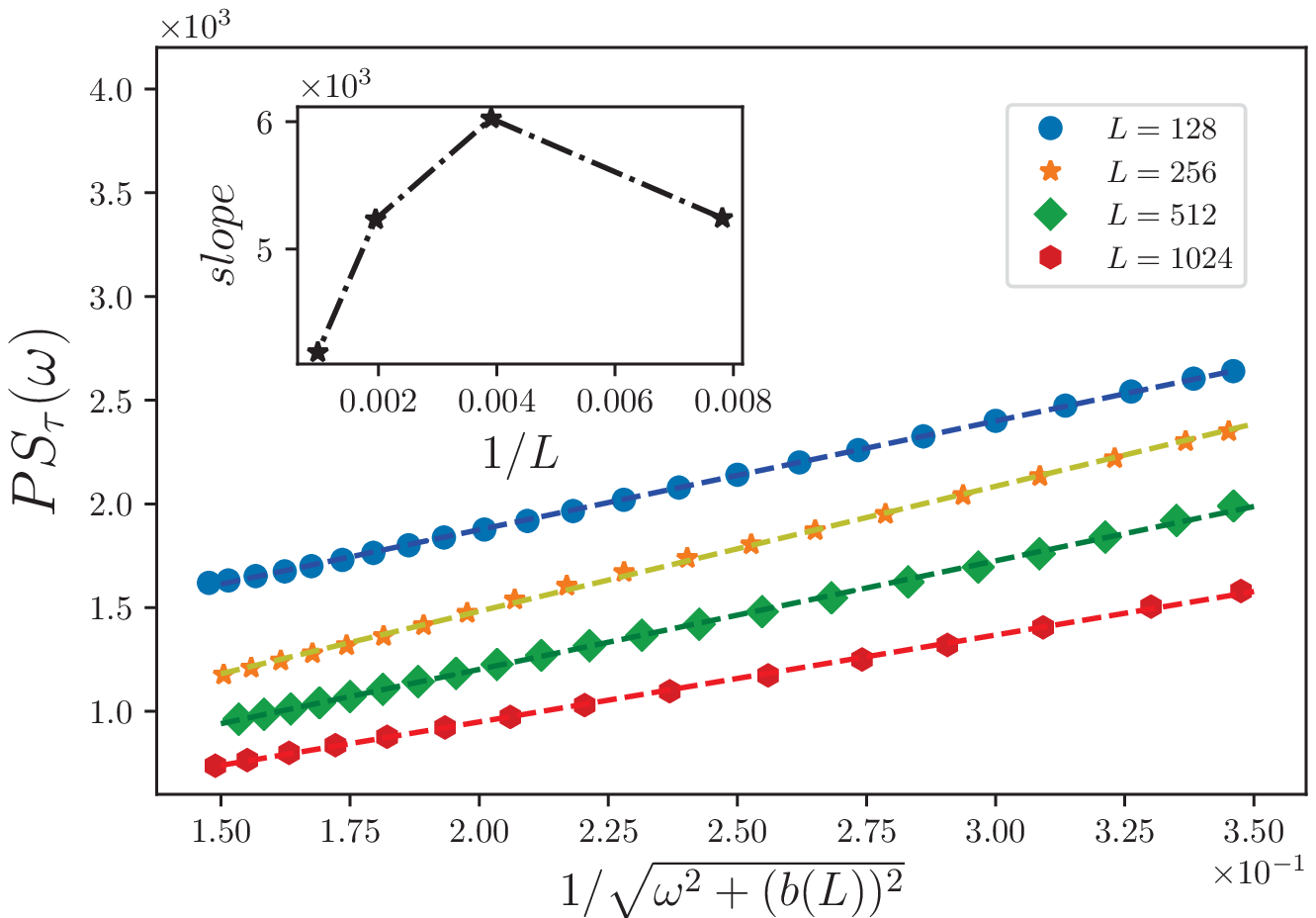}
		\caption{}
		\label{fig:PS_tau,beta}
	\end{subfigure}
	\begin{subfigure}{0.45\textwidth}\includegraphics[width=\textwidth]{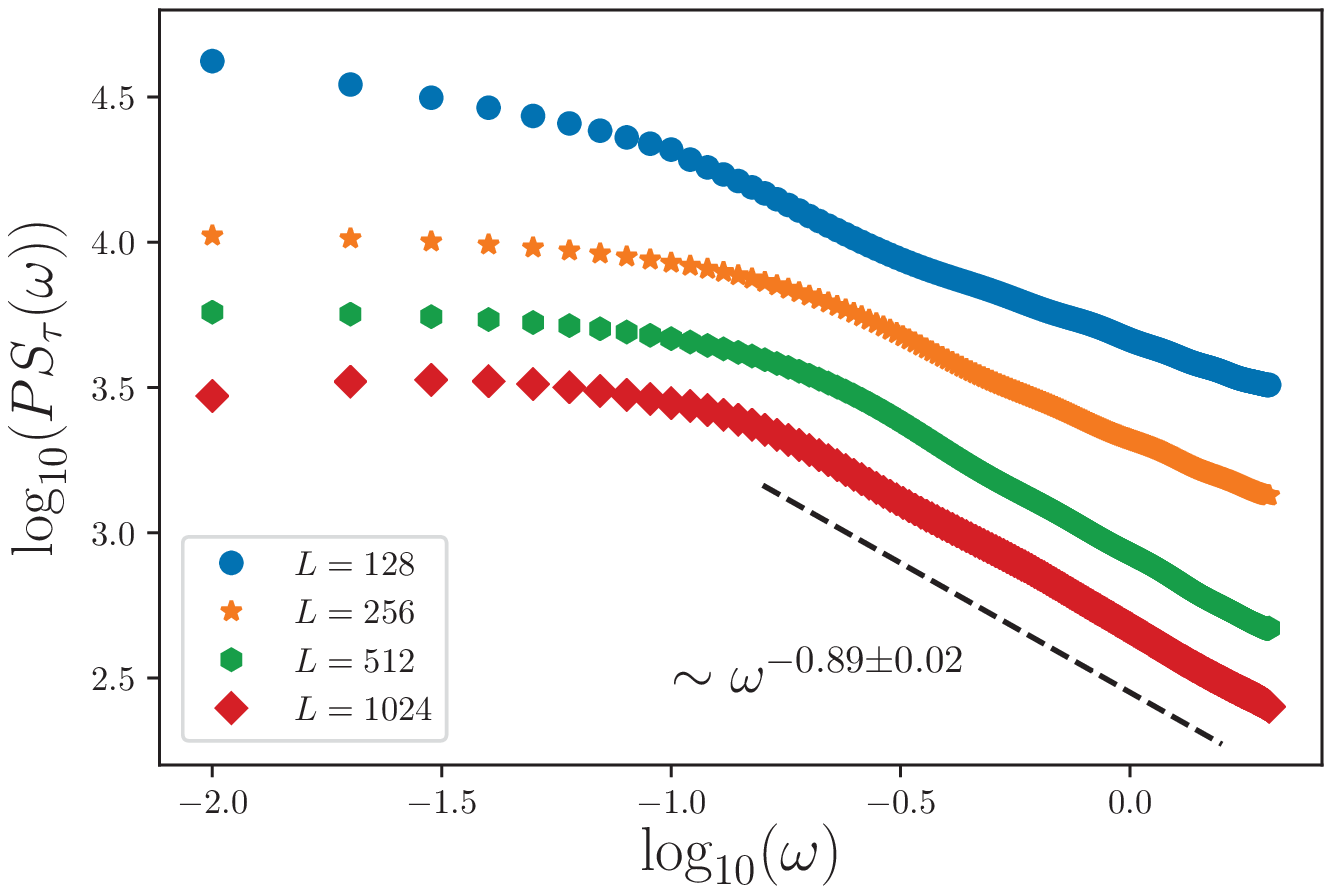}
		\caption{}
		\label{fig:PS_tau,omega}
	\end{subfigure}
	\caption{(Color online): $A_{\tau}(n)$ in terms of $n$ for various rates of lattice sizes, with the corresponding fits $A_{\tau 0}(L)\exp\left(-b_{\tau}(L)T\right)$ (broken lines). Inset: data collapse of $A_{\tau}(T)/A_{\tau 0}$ in terms of $b_{\tau}T$. (b) Log-log plot of $A_{\tau 0}(L)$ and $b_{\tau}(L)$ in terms of $1/L$. It is seen that $b_{\tau}\sim L^{-0.56\pm 0.04}$. (c) Power spectrum of $\tau$ ($PS_{\tau}(\omega)$) in terms of $\beta\equiv \frac{1}{\sqrt{\omega^2+b_{\tau}(L)^2}}$. The slop of the fitted lines has been shown in the inset in terms of $1/L$. (d) $\log_{10}PS_{\tau}(\omega)$ in terms of $\log_{10}\omega$.}
	\label{tau-statistics}
\end{figure*}

Empirically the analysis of REWT seems to be more important. The time signal $\tau(n)$ has direct connection to the statistics of the time signal $s(T)$. The statistical analysis of this quantity helps to predict vastness of the upcoming large scale events. A complete analysis of this quantity has been presented in Fig.~\ref{tau-statistics} in which all quantities presented for $S(T)$ have been regenerated. In Fig.~\ref{fig:A_tau} $A_{\tau}(n)$ has been shown, which is fitted by $-A_{0\tau}\exp\left[-b_{\tau}T\right]$ ($A_{0\tau}$ and $b_{\tau}$ are fitting parameters). The exponential behavior signals the presence of a characteristic time scale in the problem, i.e. $\delta T_{\tau}=1/b_{\tau}$ above which the correlation between the signals becomes negligibly small. The dependence of $b_{\tau}$ as well as $A_{0\tau}$ on the system size has been shown in Fig.~\ref{fig:A0_tau,b_tau}. We see that as the lattice size becomes larger, the correlations become more long range, just like $s(T)$. Having this fitting form of autocorrelations, one easily obtains that the corresponding power spectrum should be proportional to $PS_{\tau}(\omega)\sim \left(\omega^2+b_{\tau}(L)^2 \right)^{-\frac{1}{2}}$. This has been shown in the Figs.~\ref{fig:PS_tau,beta} and \ref{fig:PS_tau,omega} which display this dependence for various rates of lattice sizes. The linear dependence on $\left(b_{\tau}^2 +\omega^2\right)^{-\frac{1}{2}}$ confirms that $1/b_{\tau}$ (as $1/b_{s}$ in the $s(T)$ signal) is a new time scale in the problem. It is worthy to note here that $\delta T_{\tau}=1/b_{\tau}$ is the \textit{number of the rare event} above which the REWT signals become nearly independent, and the simple \textit{time} word should not be confused with the real times such as $T$. Here for $s(T)$ and $\tau(n)$ we see that the $1/f$ noise systematically arises for $\omega\gg b_{\tau}$ and $\omega\gg b_s$ respectively. Despite this $1/f$ noise behavior, it is very important to note that for small enough frequencies, there is a time scale which destroy the power-law behavior of the power spectrum of these functions in the BTW model.

\section*{Discussion and Conclusion}
\label{sec:conc}

Many features of the sandpile models on the various systems are known. One of the primitive aims of the BTW model had been to explain the $1/f^{\alpha}$ noise which is widely seen in the natural systems. Our work is motivated by the existence of two well-separated time scales in self-organized sandpile models, one related to the spreading of avalanches and the other imposed by the external driving. In this paper we have analyzed many features of the some noises in the BTW system. The activity inside an avalanche ($x(t)$), the avalanche sizes ($s(T)$) and the rare events waiting time (REWT) ($\tau(n)$) have been studied and the autocorrelation and the power spectrum of these noises were reported. For $x(t)$ a power-law ($1/f^{\alpha}$) decay was seen with $\alpha\simeq 1.6$ in accordance with the previous results. The multi-fractal behavior of avalanches was observed in the time dependence of $A_x(t,t')=\left\langle x(t)x(t')\right\rangle $ for which two different power-law dependence were obtained. The analysis of the probability density $P(x,t)$ shows also these distinct behaviors. For $s(T)$ and $\tau(n)$ however some exponential anti-correlation behaviors were observed. Both signals become more long-range as the lattice size increases. Despite of this fact, the amplitudes of these anti-correlations weakens as the system size increases.\\
The anti-correlation behavior of $s(T)$ shows that when a large scale event takes place, it is favorable for the system to generate smaller avalanches in the close next steps and vice versa. This behavior demonstrates that sandpiles have discharging effects. The interpretation of the anti-correlation behaviors of $\tau$'s is more complicated. This behavior means that the largeness of the time period between two successive rare events, induce a small time period between two successive rare invents in the next step and vice versa. These temporal anti-correlations have also been observed previously in the BTW model~\cite{hwa1992avalanches,ali1995self} and one-dimensional sandpile model with log-normal form autocorrelations~\cite{kutnjak1996temporal}. The spatial correlation/anticorrelation behaviors of the sandpile models can be deduced from the Hurst exponent $H$ (defined by $C_2(r)\equiv \left\langle |x(r_0+r)-x(r_0)|^2\right\rangle \sim r^{2H}$ in which $x(r)$ is the toppling number of the sand column at $t$) in which $H\simeq 0.66>0.5$ for the BTW model shows that the BTW model is smooth and less fluctuating and corresponds to a correlated surface~\cite{ahmed2010avalanche}, whereas for the model on the Bethe lattice, they are weakly anti-correlated~\cite{majumdar1991height,dhar1990abelian}. This correlation/anti-correlation behavior is model dependent in sandpile models and can be regarded as the measure of cross over between them~\cite{lubeck2000crossover}.\\
Besides the anti-correlation, the exact form of the correlation is also of central importance. The exponential dependence of the autocorrelation functions of $s(T)$ and $\tau(n)$ time signals is the result of the violation of temporal scale-invariance in the BTW model. The emergent time scales ($\delta T_s\equiv 1/b_s$ and $\delta T_{\tau}\equiv 1/b_{\tau}$) also cast the overall behavior of power spectrum to two disjoint intervals, e.g. $\omega\ll b_s$ and $\omega\gg b_s$ (and the same relations for $b_{\tau}$). For the latter case we have observed the famous $1/f$ noise, whereas for the former case the power spectrum is dominated by $b_s$. The overall dependence of the power spectrum of $s(T)$ and $\tau(n)$ are $PS_{s,\tau}(\omega)\sim\left(\omega^2+b_{s,\tau}(L)^2\right)^{-1/2}$. If our extrapolation to the large systems $L\rightarrow \infty$ is true, and $b_{s}$ and $b_{\tau}$ go to zero in a power-law fashion, then the power-spectrum $PS_{s}$ and $PS_{\tau}$ tend to the Dirac delta function in the thermodynamic limit. Therefore our results support the fact that the $1/f$ noise in the BTW model is a finite size effect.

\bibliography{refs}

\end{document}